# Properties of heavy ion beams produced by a PW sub-picosecond laser

**J. Domański[1] and J. Badziak**

*Institute of Plasma Physics and Laser Microfusion,*
*Hery Street 23, 01-497 Warsaw, Poland*
*E-mail*: jaroslaw.domanski@ifpilm.pl

ABSTRACT: The paper reports the results of numerical studies on the generation of heavy ions from a sub-micrometre gold target irradiated by a high-energy (60 J) high-intensity (~ 2 x $10^{20}$ W/cm$^2$) sub-ps laser pulse. The properties of the heavy ion beam produced from a 0.5-um gold target with and without hydrogen contaminants are investigated using a multi-dimensional (2D3V) particle-in-cell code. It was found that in the case of the pure (without the contaminants) target, the ion beam contains over 20 ion species with a charge state > 30 and a total number of high-energy (> 20 MeV) ions reaching values well above $10^{12}$. The ion energy spectrum is broad, with maximum energies > 1 GeV and mean energies > 100 MeV. At a distance ~ 1 mm from the target, the intensity of the ion beam is ~ $10^{15}$ W/cm$^2$, the ion fluence reaches ~ $10^{16}$ ions/cm$^2$, and the ion pulse duration is ~ 100 ps. The contaminants significantly reduce the ion beam parameters and both the mean ion energy and the intensity and fluence of the ion beam generated from the contaminated gold target are almost an order of magnitude lower than those produced from the pure target. Heavy ion beams with the parameters obtained in the case of the pure gold target are barely achievable in conventional RF-driven accelerators, so they can open the door to new areas of research, in particular in high energy-density physics.

---

[1] Corresponding author.

# Contents

## 1. Introduction

Recent development in high-peak-power laser technology has resulted in the construction of short-pulse lasers with petawatt powers and focused beam intensities exceeding $10^{21}$ W/cm$^2$ [1,2]. There are multi-PW lasers under construction that have anticipated beam intensities reaching up to $10^{23}$ – $10^{24}$ W/cm$^2$ [1-4]. PW and multi-PW lasers are promising drivers for charged particle acceleration, with the plasma produced by these lasers at high-intensity laser-target interaction possibly being an efficient source of high-energy electron [5,6] or ion beams [7-9]. In particular, using these lasers, heavy ion beams useful for research in nuclear physics, high energy-density physics or materials science can potentially be produced. The possibility of generating heavy ion beams from laser-produced plasma has been demonstrated in numerous experiments [10-19], but in the vast majority of these experiments the laser beam powers and/or intensities were insufficient to achieve ion beam parameters, in particular sub-GeV or GeV ion energies, desirable in most heavy ion beam applications. The use of high-energy (> 100 J) PW lasers generating sub-ps or ps pulses [1,2,20-24] to accelerate heavy ions creates the opportunity to achieve ion beam parameters useful for some applications. PW-class high-energy lasers currently operate in many countries [1,2,20-24], and examples of such lasers operating in Europe are the Vulcan PW laser in CLF RAL, Didcot, UK [20] and the PHELIX laser in GSI, Darmstadt, Germany [21].

The acceleration of heavy ions by high-energy PW-class lasers, especially ions with high mass numbers A ≥ 200, has been studied in relatively few experimental and numerical works [10,11,19,25] and the properties of very heavy ion beams produced using such lasers are still poorly understood. Comprehensive and detailed knowledge of these properties is necessary for effective programming and control of the parameters of the produced beams, which is a key condition determining the usefulness of these beams in real applications. The properties of heavy ion beams may differ significantly from those of laser-driven light ion beams due to: (1) a very high ion charge state z, (2) a high inertia of ions, (3) a large number of accelerated ion species in various charge states, and (4) a relatively low z/A ratio which is usually well below that for light ions. The low z/A ratio means that the acceleration efficiency of heavy ions is lower than that for protons or light ions, and, as a result, any light contaminants (especially hydrogen contaminants) on the surface or in the volume of the heavy ion target can significantly affect the ion acceleration process. Thus, even if the general mechanisms of acceleration of light and heavy ions are similar, the detailed run of the acceleration process for heavy ions may differ fundamentally from that of light ions.

In this paper, the properties of a heavy (Au) ion beam produced by a high-energy (60 J) sub-picosecond laser pulse from a sub-micrometre gold target with and without hydrogen contaminants are numerically investigated using a multi-dimensional (2D3V) particle-in-cell PICDOM code [26]. It was shown that, in the case of a pure (without contaminants) gold target, the laser pulse can produce more than $10^{12}$ gold ions with a mean energy > 100 MeV and a maximum energy > 1 GeV. At a distance from the target equal to ~ 1 mm, the intensity of the ion beam is ~ $10^{15}$ W/cm$^2$, the ion fluence reaches ~ $10^{16}$ ions/cm$^2$, and the ion pulse duration is ~ 100 ps. The contaminants significantly reduce the ion beam parameters; in particular, the mean ion energy and the intensity and fluence of the ion beam generated from the contaminated gold target are almost an order of magnitude lower than those produced from the pure target.

## 2. The numerical code and the laser and target parameters

The numerical investigation of the acceleration of heavy gold (Au) ions by a high-energy (60J) sub-ps laser pulse was performed using a fully electromagnetic, relativistic multidimensional (2D3V) particle-in-cell PICDOM code that includes in particular an "on-line" calculation of the ionisation level of the accelerated ions and target atoms [25,27-29], as well as the radiation losses due to synchrotron radiation produced by high-energy relativistic electrons [30,31]. The Ammasov-Delone-Krainov [32,33] formula was used to describe the ionisation process, and the Sokolov model [34] was used to describe the radiation losses. In the simulations, a thin 0.5 μm gold target was irradiated by a linearly polarised 500fs, high energy (60J) laser pulse with parameters easily achievable currently at European laser facilities such as VULCAN PW or PHELIX [1,2]. The laser pulse shape in time and space (along the y-axis) was described by a super-Gaussian function with a power index equal to 6, the laser wavelength was equal to 1.05 μm and the laser beam width was assumed to be 10 μm (these parameters correspond to a laser pulse peak intensity equal to $2.4*10^{20}$W/cm$^2$). The targets with and without a 40 nm-thick layer of hydrogen (H) contaminants situated at the back of the target were investigated. The target's transverse size was equal to 30 μm and the gold atom density corresponded to a density of solid state, equal to $5.90*10^{22}$1/cm$^3$. The density of hydrogen atoms was equal to $7.905*10^{22}$1/cm$^3$ (the value corresponds to the density of hydrogen atoms in polyethylene). Pre-plasma with a density scale length of 0.25 μm and a density shape described by an exponential function was placed in front of the target. The simulations were performed in the x,y space of dimensions 100 x 41 μm$^2$. The space steps were equal to 26 nm for a pure gold target and 8 nm for a target with contaminants, respectively, and the time step was 4.44 attoseconds. The number of ion macro-particles was assumed to be 100 Au particles/cell for the pure target, 64 Au particles/cell and 120 H particles/cell for the target with contaminants.

## 3. Results and discussion

In this paper, various characteristics of the produced heavy ion beams and the influence of hydrogen contaminants on the ion acceleration process are presented. The properties of heavy ion beams produced from a pure gold target are shown in sub-section 3.1, while the effect of hydrogen contaminants on the heavy ion beams parameters is discussed in sub-section 3.2.

**3.1. Properties of a heavy ion beam produced from a pure gold target**

The 2D spatial distribution of the density of Au ions (a) and electrons (b), as well as the electric field in the plasma along the laser beam axis $E_x$ (the field directly accelerating the ions) (c) and the ionisation level of ions (d) at the late stage of ion acceleration (simulation time t=800 fs) are presented in Fig. 1. It is visible that the velocity spectrum of ions is wide and the source of ion beam is much larger than the laser focal spot, which suggests the domination of the target normal sheath acceleration (TNSA) mechanism [8,9] in the process of ion acceleration.

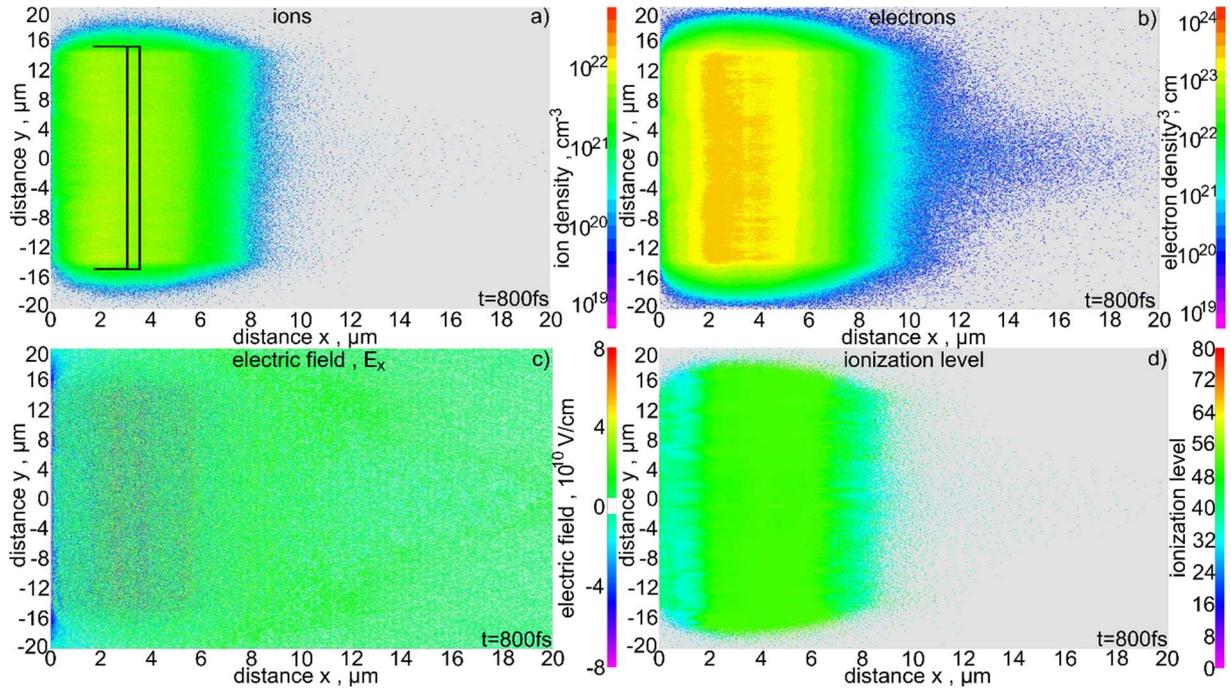

**Figure 1.** 2D spatial distribution of the density of Au ions (a) electrons (b), the electric field in the plasma along the laser beam axis $E_x$ (the field directly accelerating the ions) (c) and the ionisation level of ions (d) at the simulation time t = 800fs. The initial position of the target is indicate in (a).

Furthermore, we can observe that the distribution of the ionisation level in the ion beam is not homogenous, with the ions of the lower ionisation level dominating at the front of the beam. For a better understanding of the acceleration process, the spatial profiles of the ion and electron densities and the electric field of the accelerating ions ($E_x$) at the early and late stages of acceleration were prepared (Fig. 2). At the early stage, both the TNSA and the radiation pressure acceleration (RPA) mechanisms [8,9] contribute to the ion acceleration process, though most of the ions are accelerated by TNSA. At the late stage, the Coulomb explosion [8,18] also contributes to the process of acceleration and significantly enhance ion acceleration both in a forward and backward direction. This explosion is caused by a high charge of heavy gold ions, which is hard for electrons to screen.

The target is ionised by the laser and the TNSA field, as well as by the strong local electric field generated inside the target by hot electrons partially separated from ions. At the early stage of acceleration the target is ionised mainly close to the front and back surfaces, by the laser and TNSA fields. The influence of the "inside target" ionisation increases with time and plays a dominant role in the final stage of ionisation process. In effect, the target is ionised mostly by the strong local electric field generated inside the target. The ionisation spectrum of gold ions at the end of the ionisation process is presented in Figure 3. It extends from z = 32 to z = 54 with the distinct peak at z = 51 (Ni-like ions). The observed peaks in the spectrum correspond to the energy gaps in the ionisation energy spectrum of gold [35].

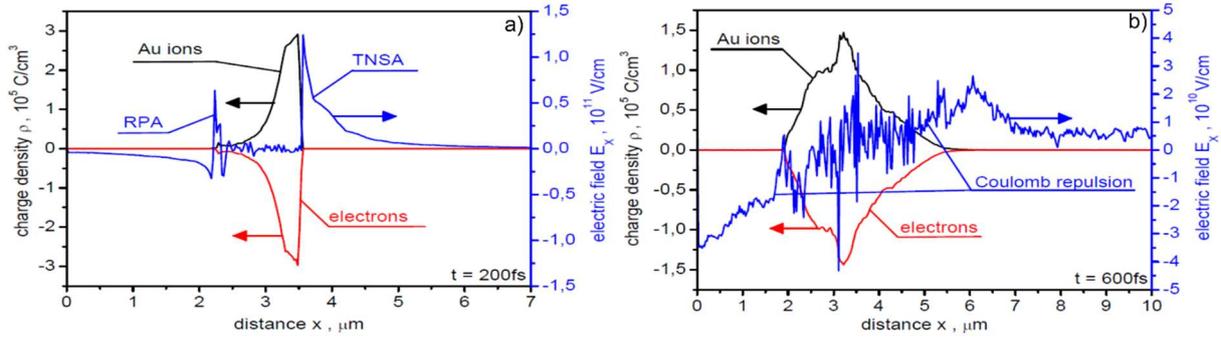

**Figure 2.** Spatial profiles of the ion density, the electron density and the electric field accelerating ions, $E_x$, along the laser beam axis at the early (a) and late (b) stages of ion acceleration.

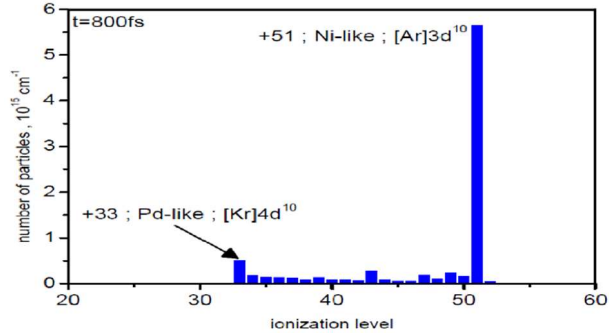

**Figure 3.** The ionisation spectrum of gold ions at the late stage of ion acceleration.

The acceleration process of ions significantly depends on its charge state (z/A ratio). The energy spectra of gold ions in various charge states (a) as well as the dependence of the mean and maximum ion energy as a function of the ion charge state (b) are presented in figure 4. Ions with different ionisation levels have essentially different energy spectra. In particular, the energy spectra of ions with $z = 43$ and $z = 33$ have a clearly non-Maxwellian shape, while the energy spectrum of ions with $z = 51$ is quasi-Maxwellian. Both the mean and the maximum ion energies depend substantially on the ion charge state. Unexpectedly, the mean energy of ions decreases with increasing z, while the highest maximum ion energy is observed at $z = 43$. The mean ion energies cover the range from 34 MeV to 200 MeV, while the maximum ion energy reaches values even above 1GeV. However, the number of ions with such high energies is very low. Additionally, Figure 4c presents the total energy of gold ions and the laser-to-ions energy conversion efficiency as a function of the ion charge state. It is visible that most of the ion beam energy is stored in $Au^{+51}$ ions, which form a distinct peak in the ionisation spectrum shown in Figure 3. The observed dependence of mean ion energy on ionization level could be explained by strength and type of the field acting on ions of different ionisation level. The ions of lower ionisation level are accelerated by the TNSA field during the first part of the acceleration process. In effect, they leave the area of target and strong field, and the process of ionisation is stopped. In opposite most of particles with lower velocities remain closer to their original position and are ionised by "inside target" field. However the complicated structure of the electric field in this region, especially in the late stage of acceleration, leads to the non-effective acceleration process in part of ions and lower mean ion energy. The difference in the energy spectra also could be explain by the above mentioned effect. Only ions of enough high velocity could leave the area of the target during ionisation process. The rest of the ions remain close to their origin position and are ionised to the higher level. For this reason a clearly non-Maxwellian shape of the energy spectra for low-z types of ions are observed.

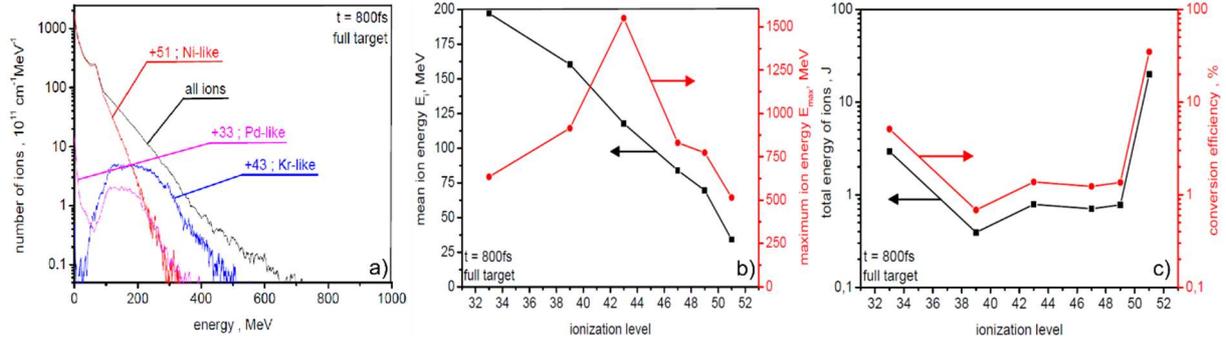

**Figure 4.** The energy spectra of gold ions in various charge states (a) and the mean and maximum ion energy (b) as well as the total energy of gold ions and the laser-to-ions energy conversion efficiency (c) as a function of the ion charge state.

From the point of view of the potential applications of ion beams, parameters such as the ion beam intensity, the temporal shape, the fluence or the total number of accelerated ions are important. Figure 5 presents the temporal shape of gold ion beam generated from the investigated target and recorder at a short distance (a, $l_d$ = 10 µm) and a long distance (b, $l_d$ = 1 mm) behind the target. In addition, the ion pulse shapes for ions in selected charge states are also shown in Figure 5a. The transverse size of the beam was calculated assuming an angular divergence of the ion beam equal to 10 degrees (recorded in the simulation) and the size of the ion source was equal to the transverse size of the target. Furthermore, the correction resulting from the three-dimensionality of the beam (3D correction) was taken into account. The ion pulse shapes are similar for two considered positions of the detector, while the ion pulse duration increases nearly linearly with the distance from the target (from ~ 1 ps to ~ 100 ps) due to the ion velocity dispersion.

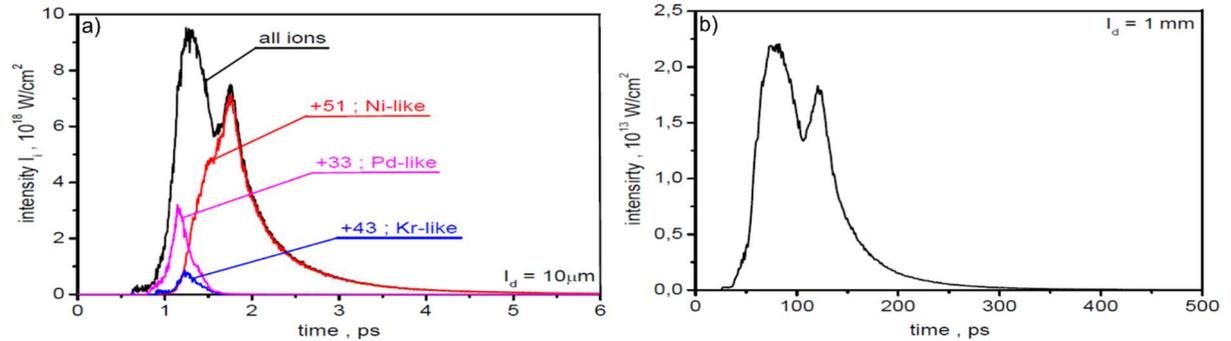

**Figure 5.** The temporal shape of gold ion beam, for all ions and for ions in selected charge states, recorded at a short distance (a, $l_d$ = 10µm) and a long distance (b , $l_d$ = 1mm) behind the target.

The ion pulse intensity decreases quickly with distance, and at a distance of 1 mm from the target it is more than 3 orders of magnitude lower than at the ion source. Both the ion velocity dispersion and the ion beam angular divergence are the reason for the observed rapid decrease in pulse intensity. However, even at relatively long distance from the target (~ mm), the ion pulse intensity is very high (~ $10^{15}$ W/cm$^2$), several orders of magnitude higher than achieved in RF-driven heavy ion accelerators.

Table 1 shows the values of ion fluence for different ranges of ion energy at a distance of 10 µm and 1 mm from the target, while the number of gold ions for different ranges of ion energy for ions generated in a forward direction are presented in Table 2. The number of high-energy (> 20 MeV) ions reaches values well above $10^{12}$ ions/shot, and the ion fluence at a distance of 1 mm is equal to ~ $10^{16}$ ions/cm$^2$ (much higher than in RF-driven heavy ion accelerators).

| fluence $F_i$, $10^{16}$ 1/cm² | all ions | E>1 MeV | E>10 MeV | E> 20MeV |
|---|---|---|---|---|
| detector position = 10 µm | 113.09 | 109.14 | 83.36 | 66.78 |
| detector position = 1 mm | 1.60 | 1.58 | 1.44 | **1.29** |

**Table 1.** The ion fluence for different ranges of ion energy at a distance of 10 µm and 1 mm from the target.

| number of ions, $10^{12}$ | all ions | E>1 MeV | E>10 MeV | E> 20MeV |
|---|---|---|---|---|
|  | 10.3 | 9.69 | 7.16 | **5.70** |

**Table 2.** The number of gold ions for different ranges of ion energy for ions generated in a forward direction.

### 3.2. The effect of hydrogen contaminants on the heavy ion beam parameters

The light atom contaminants naturally present at the surfaces of the targets could have an important influence on the process of heavy ion acceleration [25,27]. For this reason, a comparison of results from the simulations performed for the pure gold target (see the previous sub-section) and the gold target with a 40 nm thick layer of hydrogen contaminants was performed. Figure 6 presents the spatial profiles of the ion density, the electron density and the electric field accelerating ions, $E_x$, along the laser beam axis at the early stages of ion acceleration and for the target without (a) and with (b) contaminants. It can be seen that the presence of contaminants reduces the strength of the TNSA field by a factor of ~ 4. In addition, the peak of this field is detached from the front part of the gold ion beam. This effect could be explain by the screening of the TNSA field by protons accelerated from contaminants.

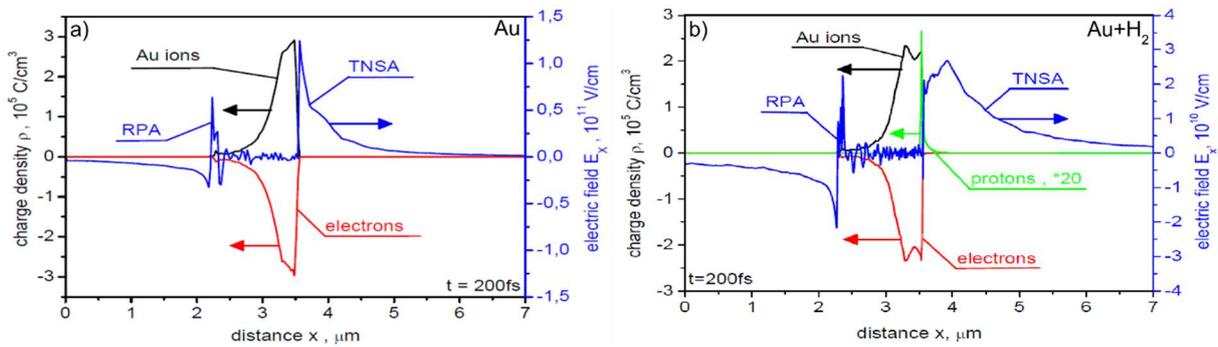

**Figure 6.** Spatial profiles of the ion density, the electron density and the electric field accelerating ions, $E_x$, along the laser beam axis at the early stages of ion acceleration. Target without (a) and with (b) contaminants.

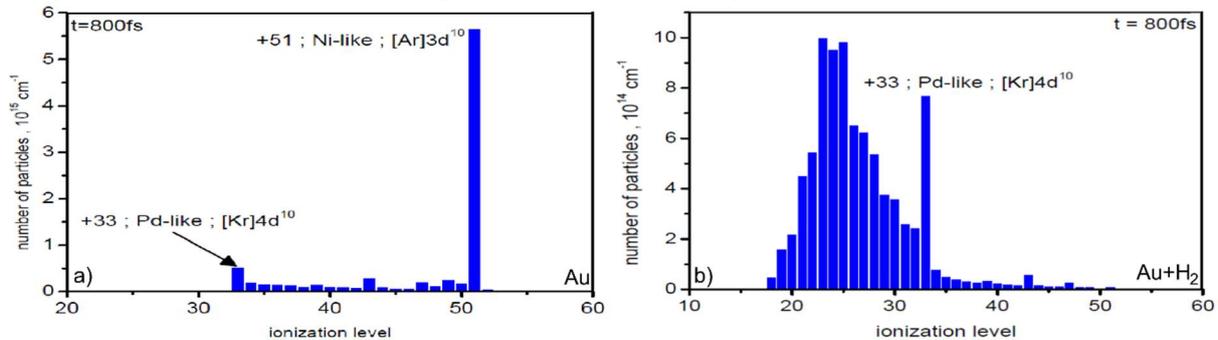

**Figure 7.** The ionisation spectra of gold ions at the late stage of ion acceleration. Target without (a) and with (b) contaminants.

Furthermore, the strong local electric field generated inside the target by hot electrons partially separated from ions is significant reduced. TNSA and "inside target" fields several times lower acting on ions leads to huge changes in the process of target ionisation and the final ionisation spectrum of the gold ions (Figure 7). The presence of contaminants significantly reduces the ionisation level of the accelerated gold ions: the mean ionisation level is reduced from around 50 to around 25-30. The reduction in the accelerating field and the ionisation level have a dramatic negative impact on the final energy of the accelerated gold ions. Figure 8 presents the mean (a) and maximum (b) ion energy as a function of time for the target without and with contaminants. It can be seen that, at the end of the simulation, the mean ion energy for the target with contaminants is 7 times smaller than for the target without contaminants. The maximum ion energies are similar for both investigated cases.

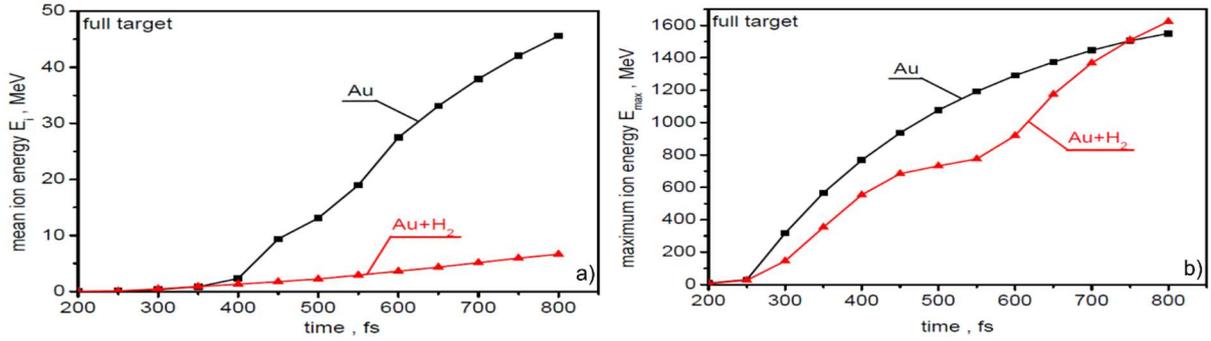

**Figure 8.** The mean (a) and maximum (b) ion energy as a function of time for the target with and without contaminants.

The observed decrease in the mean energy of gold ions has a significant influence on other parameters of the gold ion beam, such as the beam intensity, temporal shape of the beam or ion fluence. The temporal shape of the gold ion beam (a) and the ion fluence (b) for forward-generated high energy (> 20 MeV) gold ions as a function of time for both types of investigated cases are presented in Figure 9. We can see that the duration and intensity of the gold ion pulse for the target with contaminants are several times smaller than for the target without contaminants (Fig. 9a), while the ion fluence is even more than an order of magnitude lower for the target with contaminants (Fig. 9b). To conclude, hydrogen contaminants have a significant negative influence on the process of heavy ion acceleration, and considerably reduce almost all of the parameters of the heavy ion beam.

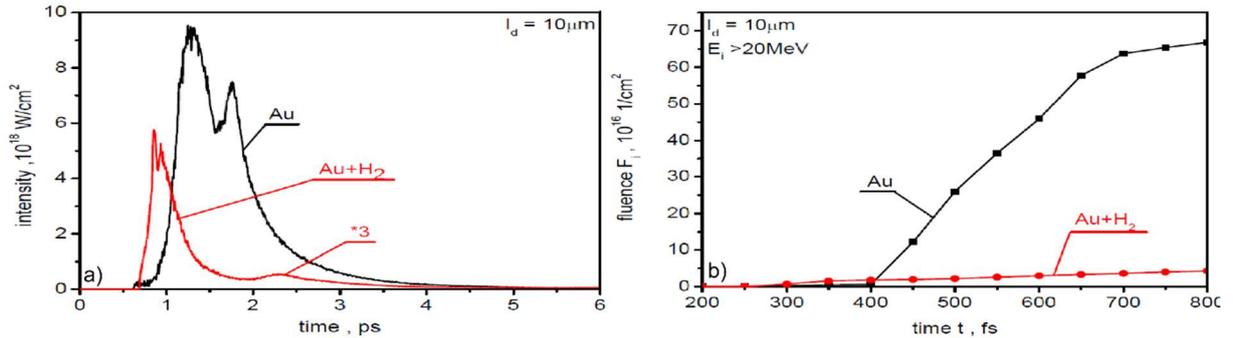

**Figure 9.** The temporal shape of the gold ion beam (a) and the ion fluence for high energy (> 20 MeV) gold ions as a function of time (b) recorded 10 μm behind the target with and without contaminants. For better visualisation the values of ion beam intensities for Au-$H_2$ target are multiply by 3.

## 3. Conclusions

The paper reports the results of numerical studies on the acceleration of heavy (Au) ions from a 0.5-µm gold target irradiated by a sub-PW sub-ps laser pulse with parameters easily achievable currently at European laser facilities such as VULCAN PW or PHELIX.

It has been found that, for a pure (without contaminants) target, the ionisation spectrum extends from $z = 32$ to $z = 54$, with a distinct peak at $z = 51$, and that most of the energy of the ion beam is stored in $Au^{+51}$ ions. The ions in different charge states have essentially different energy spectra. In particular, the energy spectrum of ions with $z = 43$ have a clearly non-Maxwellian shape and mean ion energy equal to 200 MeV, while the energy spectrum of ions with $z = 51$ is quasi-Maxwellian, with mean ion energy equal to 34 MeV. At a distance ~ 1 mm from the target, the intensity of the ion beam is very high ( ~ $10^{15}$ W/cm$^2$), the ion fluence reaches ~ $10^{16}$ ions/cm$^2$, and the duration of the ion pulse is ~ 100 ps. Furthermore, the number of high-energy (> 20 MeV) ions reaches values well above $10^{12}$ ions/shot. Heavy ion beams with such parameters are barely achievable in currently operating RF-driven accelerators, so they can open the door to new areas of research, in particular in high energy-density physics.

Hydrogen contaminants have a significant influence on the process of heavy ion acceleration. In particular, the presence of the contaminants decreased the ionisation level of accelerated ions and reduced ion beam parameters such as the mean ion energy, the ion beam intensity and the beam fluence by about an order of magnitude. For this reason, the decontamination of heavy ion targets before starting the laser driver-target interaction is of crucial importance.

The thickness of the pre-plasma layer (laser pulse contrast ratio) can also have a significant impact on the parameters of the generated heavy ion beam. Because the laser pulse interacts mainly with this layer, changing its thickness will affect not only the RPA mechanism of ion acceleration, dominant at the front of the target, but also the amount and temperature of hot electrons generated by the laser, and thus the efficiency of the TNSA mechanism and the value of the electric field inside the target. Obtaining high ion beam parameters therefore requires careful control of pre-plasma parameters.

## Acknowledgments

The simulations were carried out with the support of the Interdisciplinary Center for Mathematical and Computational Modelling (ICM), University of Warsaw under grant no. G57-20 and the Poznan Supercomputing and Networking Centre under grant no. 417.